\documentstyle[12pt,psfig]{article}
\def\be{\begin{equation}}
\def\ee{\end{equation}}

\def\TL{\hfil$\displaystyle{##}$}
\def\TR{$\displaystyle{{}##}$\hfil}

\def\TT{\hbox{##}}
\def\seqalign#1#2{\vcenter{\openup1\jot
  \halign{\strut #1\cr #2 \cr}}}

\def\fixit#1{}

\def\mop#1{\mathop{\rm #1}\nolimits}

\def\Tr{\mop{Tr}}

\def\href#1#2{#2}

\def\eqalign#1{\vcenter{\openup1\jot
    \halign{\strut\span\TL & \span\TR\cr #1 \cr
   }}}
\def\lbldef#1#2{\expandafter\gdef\csname #1\endcsname {#2}}

\begin{document}
\baselineskip=16pt
\pagestyle{plain}
\setcounter{page}{1}

\begin{titlepage}

\begin{flushright}
hep-th/0304079 \\
PUPT-2080
\end{flushright}
\vfil

\begin{center}
{\huge Proton Decay in Intersecting D-brane Models}
\end{center}

\vfil
\begin{center}
{\large
Igor R. Klebanov\footnote{On leave from Princeton University.}
and
Edward Witten
}\end{center}

$$\seqalign{\span\TL\, & \sl\span\TT}{
 & Institute for Advanced Study,
 Princeton, NJ  08540, USA  \cr
}$$
\vfil

\begin{center}
{\large Abstract}
\end{center}

We analyze proton decay via dimension six operators in certain
GUT-like models derived from Type IIA orientifolds with
$D6$-branes.  The amplitude is parametrically enhanced by a factor
of $\alpha_{GUT}^{-1/3}$ relative to the coresponding result  in
four-dimensional GUT's. Nonetheless, even assuming a plausible
enhancement from the threshold corrections, we find little overall
enhancement of the proton decay rate from dimension six operators,
so that the predicted lifetime from this mechanism remains close
to $10^{36}$ years.

\vfil
\begin{flushleft}
April, 2003
\end{flushleft}
\end{titlepage}
\newpage
\renewcommand{\thefootnote}{\arabic{footnote}}
\setcounter{footnote}{0}
\renewcommand{\baselinestretch}{1.2}  
\tableofcontents
\def\bar{\overline}

\section{Introduction}
\label{Introduction}

Grand Unified Theories in four dimensions have had impressive
successes, in accounting for the quantum numbers of fermions
\cite{gg}, in predicting -- with the aid of supersymmetry -- a
value of $\sin^2\theta_W$ that is in excellent agreement with
experiment \cite{raby}, and in pointing twenty years in advance of
decisive measurements to the right order of magnitude of neutrino
masses \cite{seesaw}. GUT theories also make the exciting
prediction that the proton may decay with a lifetime close to
present experimental bounds.

With supersymmetry, even if one assumes an $R$-parity symmetry to
avoid catastrophic proton decay at low energies, there actually
are two different GUT-based mechanisms for proton decay.  The
proton may decay due to dimension five operators of the form $\int
d^2\theta Q^3L$, where $Q$ and $L$ are quark and lepton
superfields.  Or it may decay by dimension six operators of the
form $\int d^4\theta Q^2 \tilde Q^*\tilde L^*$ coming from gauge
boson exchange; this effect is similar to proton decay in GUT's
without supersymmetry, but is significantly slower because
supersymmetry raises the GUT scale. In the simplest models, proton
decay by dimension five operators dominates, and in fact present
experimental bounds make life difficult for these models
\cite{murayama, hisano, rabynew} (see, however, \cite{Bajc}). This suggests the possibility
that some mechanism suppresses the dimension five operators and
that gauge boson exchange may after all  be the dominant
mechanism.  Various methods to suppress the dimension five
operators while preserving the successes of GUT's are known,
though they tend to be somewhat technical.  For example, one
construction based on discrete symmetries \cite{witten} is fairly
natural in the class of models that we will consider in this
paper.

Most GUT-like string-based models of particle physics do not
precisely lead to four-dimensional GUT's, since unification takes
place only in higher dimensions (for an early review, see
\cite{gsw}, chapters 15 and 16), leading among other things to
possibilities for GUT symmetry breaking by discrete Wilson lines
\cite{chsw}  and to a higher-dimensional mechanism for
doublet-triplet splitting \cite{oldwitten}.  Generally, models
such as the heterotic string on a Calabi-Yau manifold lead to
qualitatively similar issues concerning proton decay to those in
GUT's, though the details are somewhat different.  One important
reason that the details are different is that because of the
higher-dimensional structure, there is generally in these models
no precise answer to the question, ``Which color triplet gauge
boson is the $SU(5)$ partner of the standard model gauge bosons?''
There is a lightest color triplet gauge boson, whose wave-function
in the compact dimensions depends on the details of the model, and
its exchange may, depending on the model, give the right order of
magnitude though not the correct numerical value of the dimension
six part of the proton decay amplitude.

Recently \cite{FW}, these issues were reconsidered in another
class of models -- $M$-theory on a $G_2$ manifold.  (Actually, in
many cases these models give dual descriptions, useful in a
different region of the parameter space, of the same models that
can be studied via the heterotic string on a Calabi-Yau manifold.)
The GUT threshold correction to proton decay was computed in this
class of models and was seen to give a potential enhancement of
the proton decay rate.  It was also shown that in this type of
model, because of the way quarks and leptons are localized,
exchange of the lightest color triplet gauge boson does not
dominate the proton decay amplitude.  On the contrary, a field
theoretic attempt to compute a proton decay amplitude by summing
over Kaluza-Klein harmonics runs into an ultraviolet divergence,
and consequently the correct answer depends on the cutoff provided
by $M$-theory. Formally, this UV divergence enhances the proton
decay amplitude by a factor of $\alpha_{GUT}^{-1/3}$ compared to
what it would be in a four-dimensional model. (This factor must be
combined with the threshold factor, of course.) The coefficient of
$\alpha_{GUT}^{-1/3}$ is universal and independent of the details
of the model (such as how the GUT symmetry is broken) -- in fact,
the dominant four-fermi operator is local and invariant under the
full GUT symmetry, in contrast to the usual situation in
four-dimensional GUT's.  However, present knowledge of $M$-theory
does not make it possible to compute the numerical coefficient of
this operator.

As an alternative, we will in the present paper consider another
dual class of models in which the calculation can be done. These
models fall in the general class of intersecting brane worlds
\cite{berk,Blumenhagen:2000wh,Angelantonj:2000hi,Aldazabal,Blumenhagen:2000ea,Blumenhagen:2001te,Kokorelis} 
-- Type IIA superstrings with gauge bosons
supported on $D6$-branes and chiral matter multiplets at
intersections of the $D6$-branes. Many supersymmetric
GUT-like models have been constructed along these lines
\cite{Cvet,Blumenhagen:2002gw,CvetPap}, 
though so far none with the mechanism of symmetry
breaking by Wilson lines on the $D6$-brane that we will assume in
the present paper to justify borrowing various above-cited
results.\footnote{
In the dual context of M-theory on $G_2$ manifolds,
some examples of
symmetry breaking by Wilson lines were considered 
in \cite{tf}.} 
(In actual model-building, one typically considers branes
in an orbifold of a six-torus, but more generally one may start
with any Calabi-Yau manifold: see, for example, \cite{Blumenhagen:2002wn}.) Threshold corrections in intersecting
D6-brane GUT-like models were calculated in \cite{Stieb}. 
These models are dual to $M$-theory
on a manifold of $G_2$ holonomy as $g_{s}$, the Type IIA string
coupling constant, becomes large.  The chiral matter fields are
localized (at brane intersections) in a similar way to what
happens in $M$-theory on a $G_2$ manifold, and accordingly we will
find the same anomalous factor of $\alpha_{GUT}^{-1/3}$ in the
proton decay amplitude. The difference is that in perturbative
Type IIA string theory, everything is explicitly calculable, and
hence we will be able, at least for $g_s\ll 1$, to be precise
about the numerical factors. Regrettably, the numerical factors
accompanying $\alpha_{GUT}^{-1/3}$ are such that even assuming a
plausible enhancement from the threshold factors, the proton decay
rate is comparable to or only slightly greater than that in
standard SUSY-GUT's.
  Of course, the precise factors that appear in the corresponding
$M$-theory (or large $g_s$) limit are still unknown.

In this paper we will not be concerned with any specific model,
but will rather try to incorporate universal features of the
GUT-like intersecting D-brane models relevant to calculation of
the proton decay rate. Typically one needs a stack of D6-branes
intersecting an orientifold fixed sixplane along $3+1$ dimensions.
On the covering space we then have a stack of D6-branes
intersecting an image set of D$6'$-branes. If there are 5
D6-branes in the stack, then on the covering space we find gauge
group $SU(5)\times SU(5)$, with open strings localized at the
intersection transforming in the bifundamental representations
$({\bf 5}, {\bf {\overline 5}})+ ({\bf {\overline 5}}, {\bf 5})$.
After the orientifold projection we find $SU(5)$ gauge theory with
matter in ${\bf 10}+ {\bf {\overline {10}}}$. Our goal is to
calculate the 4-fermion contact term whose $SU(5)$ structure is
${\bf 10}^2{\bf{\overline {10}}}{}^2$. Such an interaction
mediates proton decay processes such as $p\to \pi^0e^+_L$ (and,
depending on the assumed flavor structure, other modes with
$e^+_L$ replaced by $\mu^+_L$ and/or $\pi^0$ by $K^0$). The
calculation is conveniently carried out on the covering space
where we need the 4-point function for two $({\bf 5}, {\bf
{\overline 5}})$ states and two $({\bf {\overline 5}}, {\bf 5})$
states. The calculation is  sensitive to the local structure of
the intersection, and is  insensitive to how the D6-branes are
wrapped around the compact space, as long as its size is somewhat
greater than the string scale so that worldsheet instantons are
suppressed.

Four-dimensional GUT's also have dimension six operators ${\bf
10}\,\bf {\overline {10}}\,{\bf 5}\,{\bf\overline 5}$ which lead
to proton decay modes such as $p\to\pi^0 e^+_R$ and
$p\to\pi^+\nu$; as observed in \cite{FW} in the analogous
$M$-theory case, such interactions do not arise generically in a
model of this type.  To get such an interaction, we need two
stacks of branes that are not mirror images to meet on an
orientifold plane; in general, this would be a coincidence. Both
four-dimensional GUT's and the brane worlds have additional ${\bf
5}^2 {\bf {\bar 5}}{}^2$ interactions (which in the brane worlds arise
from brane intersections away from the orientifold planes), but
these do not violate baryon number.

\section{Vertex Operators}

Without a loss of generality we may assume that
the D6-branes are oriented in the $0123468$ directions.
The D$6'$-branes intersect them along the $0123$ directions
that will have the interpretation of a $3+1$ dimensional
`intersecting brane world.' To specify its orientation in the six
transverse directions, we define the complex coordinates
\begin{equation}
z_1= x^4 + i x^5\ , \quad
z_2= x^6 + i x^7\ , \quad
z_3= x^8 + i x^9\ .
\end{equation}
In order to preserve ${\cal N}=1$ supersymmetry in $3+1$
dimensions, the rotation must act as an $SU(3)$ matrix on the
three complex coordinates \cite{berk}. Choosing the matrix to be
diagonal, we see that the rotation that turns the D6-branes into
the D$6'$-brane acts as
\begin{equation}
z_1\rightarrow \exp( \pi i \theta_1) z_1\ , \qquad
z_2\rightarrow \exp( \pi i \theta_2) z_2\ , \qquad
z_3\rightarrow \exp( \pi i \theta_3) z_3\ ,
\end{equation}
where
\begin{equation}
\theta_1+ \theta_2+ \theta_3 = 2 \quad {\rm mod~~ 2{\bf Z}}
\ .
\end{equation}

We need to construct the vertex operators for the $6-6'$
and $6'-6$ open
strings. Such operators, when inserted on the boundary of the
disk, create discontinuities in the boundary conditions for the
transverse coordinates and their fermionic partners \cite{Aki}. A standard
method for studying correlators on the upper half-plane is the
doubling trick where the half-plane is replaced by the entire
plane, but with only the holomorphic part of the field on it (for a review, see \cite{HK}). If
we consider complex fields $X^i$ corresponding to the transverse
coordinates $z^i$, then their Laurent expansion around the
insertion of the vertex operator has mode numbers shifted by
$\theta_i$ \cite{Polch}. This is analogous to what happens near a
twist field introduced in orbifold theories and we can use similar
methods for calculating the correlation functions.

Consider a bosonic twist field $\sigma_+$ which creates a
discontinuity in a complex coordinate $X$.
Then we have the OPE \cite{dfms,hv,br}
\begin{equation}
\partial X (z) \sigma_+ (w) \sim
(z-w)^{\theta-1} \tau_+\ , \qquad
\partial \bar X (z) \sigma_+ (w) \sim
(z-w)^{- \theta} \tau_+'\ .
\end{equation}
The dimension of $\sigma_+$ is $\Delta_\sigma=\theta (1- \theta)/2$,
and we will assume that $\theta$ lies between $0$ and $1$.
Now consider the complex fermion $\psi$, which is a
worldsheet superpartner of $X$. By worldsheet supersymmetry,
its mode numbers are also shifted by $\theta$.
In the Ramond sector, $\sigma_+$ must be accompanied by the
fermionic twist $s_+$ such that
\begin{equation}
\psi (z) s_+ (w) \sim
(z-w)^{\theta- 1/2} t_+\ , \qquad
\bar \psi (z) s_+ (w) \sim
(z-w)^{1/2- \theta} t_+'\ .
\end{equation}
If we bosonize $\psi$ into a real scalar $H$,
whose Green's function is
\begin{equation}
\langle H(z) H(w) \rangle =- \log (z-w)
\ ,
\end{equation}
 then
\begin{equation}
\psi= \exp (iH)\ ,\quad
\bar \psi= \exp(-iH) \ ,\quad
s_+ =
\exp [i(\theta - 1/2) H]\ .
\end{equation}
This means that the dimension of $s_+$ is
\begin{equation}
\Delta_s = {1\over 2}\left (\theta- {1\over 2} \right )^2
\ .
\end{equation}
Therefore,
\begin{equation}
\Delta_s + \Delta_\sigma= 1/8
\ ,
\end{equation}
independent of the rotation angle $\theta$.

The complete vertex operator that creates a massless $6-6'$ string in
the R sector is
\begin{equation}
V_+ = \lambda^i_{\ I} u_\alpha
e^{-\phi/2} S^{\alpha} e^{ik_\mu X^\mu}
\prod_{i=1}^3 (\sigma_+^i s_+^i)
\end{equation}
where $\phi$ is the bosonized ghost field.
If we bosonize the fermions $\psi^\mu$, $\mu=0,1,2,3$, into scalars
$h_1$ and $h_2$, then the $3+1$ dimensional spinor
$S^\alpha
\sim \exp [i (s_1 h_1+ s_2 h_2)/2] $, where $s_1,s_2=\pm
1$. The GSO projection for $6-6'$ strings requires that $s_1 s_2
=1$, which restricts us to a spinor of definite chirality.
$\lambda^i_I$ is a Chan-Paton factor with an index $i$ in the
fundamental of the first $SU(5)$, and an index $I$ in the
antifundamental of the second $SU(5)$. Since $h( e^{-\phi/2} )=
3/8$ and $h (S^\alpha )= 1/4$, we see that the requirement that
the dimension of $V_+$ is $1$ gives the 4-d massless dispersion
relation $k^2=0$.

The vertex operator $V_-$
for $6'-6$ strings is constructed similarly,
except we need to replace the $\sigma_+$ and $s_+$ twists by
$\sigma_-$ and $s_-$. The latter are defined by
sending $\theta \rightarrow 1-\theta$ in the OPE.
The GSO projection now requires that $s_1 s_2 =-1$;
this corresponds
to 4-d spinor chirality opposite to that of the $6-6'$ string.
Thus, we have the vertex operator
\begin{equation}
V_- = \tilde \lambda^I_{\ i} \bar u_{\dot \alpha}
e^{-\phi/2} S^{\dot \alpha} e^{ik_\mu X^\mu}
\prod_{i=1}^3 (\sigma_-^i s_-^i)
\ .\end{equation}

\section{The 4-fermion Amplitude in String Theory}

Our goal is to calculate the 4-fermion amplitude
\begin{equation}
\int_0^1 dx \langle V_- (0) V_+ (x) V_- (1) V_+ (\infty) \rangle
\ .
\end{equation}
The crucial ingredient in this calculation is the knowledge of the
correlator of the bosonic twist fields $\sigma$. In the orbifold
theories, these were calculated in \cite{dfms, hv, br}. We will
take the square root of their result to account for the fact that
we have the boundary twist fields rather than the bulk ones. (This
intuitively plausible result has been derived in
\cite{Cvetic} and, up to normalization, in \cite{Abel}; see also \cite{Gava,Anton} for a treatment of cases
where the intersecting D-branes have different dimensionalities.)
This gives
\begin{equation}
\langle \sigma_- (0) \sigma_+ (x) \sigma_- (1)
\sigma_+ (\infty) \rangle \sim \sqrt{\sin (\pi \theta)}
[ x (1-x)]^{-2 \Delta_\sigma} [F(x) F(1-x)]^{-1/2}
\ ,
\end{equation}
where
\begin{equation}
F(x)\equiv F(\theta,1-\theta; 1; x)
\end{equation}
is the hypergeometric function and $\sim$ represents a numerical
constant.

On the other hand, the correlator of the fermionic twists is simply
\begin{equation}
\langle s_- (0) s_+ (x) s_- (1)
s_+ (\infty)\rangle \sim
[ x (1-x)]^{-2 \Delta_s}
\ ,
\end{equation}
while the ghost factor contributes
\begin{equation}
\langle e^{-\phi/2} (0) e^{-\phi/2} (x) e^{-\phi/2} (1)
e^{-\phi/2} (\infty)\rangle \sim
[ x (1-x)]^{-1/4}
\ .
\end{equation}
The correlator of the 4-d spin fields multiplied by
spinor polarizations gives
an $x$-independent factor
$\bar u_1 \gamma^\mu u_2 \bar u_3 \gamma_\mu u_4 $.
This $x$-independence is evident from the form of the bosonized
correlator
\begin{equation}
\langle \exp [i (h_1- h_2)/2] (0) \exp [i (h_1+ h_2)/2] (x)
\exp [- i (h_1- h_2)/2] (1)
\exp [-i (h_1+ h_2)/2] (\infty)\rangle
\ .
\end{equation}

Putting together all the parts of the vertex operators, defining
as usual
\begin{equation}
s=- (k_1+ k_2)^2\ ,\qquad
t=- (k_2+ k_3)^2
\ ,
\end{equation}
and including both $s$- and $t$-channel diagrams, we find the
amplitude
\begin{equation}  \label{ampl}
\eqalign{A_4 &(k_1,k_2 ,k_3,k_4)=i C (2\pi)^4 \delta^4
(\sum_{i=1}^4 k_i) \bar u_1 \gamma^\mu u_2 \bar u_3 \gamma_\mu u_4
\ \Tr\left(\tilde\lambda_1 \lambda_2 \tilde\lambda_3
\lambda_4+\tilde\lambda_1 \lambda_4 \tilde\lambda_3 \lambda_2\right) \cr
&\cdot \int_0^1 dx x^{-1-\alpha' s} (1-x)^{-1-\alpha' t}
\prod_{i=1}^3 \sqrt{\sin (\pi \theta_i)} [ F(\theta_i ,
1-\theta_i; 1; x) F(\theta_i , 1-\theta_i; 1; 1-x) ]^{-1/2} \
,\cr}
\end{equation}
where $C$ is a normalization that we will fix
later, while
the $\lambda$'s and $\tilde\lambda$'s are Chan-Paton
matrices of $6-6'$ and $6'-6$ strings.
In four dimensions, with
$\bar u_1,\bar u_3$ having one chirality and $u_2,u_4$ the other, we have
$\bar u_1 \gamma_\mu u_2\bar u_3\gamma^\mu u_4
= -\bar u_1 \gamma_\mu u_4\bar u_3\gamma^\mu u_2 $. 
This insures the antisymmetry of the amplitude under
permutation of particles $2$ and $4$ (or of particles $1$ and $3$).

We note that the hypergeometric functions are crucial for
the convergence of the integral at the endpoints of the
integration region for vanishing momenta $k_i$.
As $x\rightarrow 0$,
\begin{equation} \label{asymptotics}
F(x) \rightarrow 1\ ,\qquad
F(1-x) \rightarrow {\sin (\pi\theta)\over \pi} \ln (\delta/x)
\ ,
\end{equation}
where
\begin{equation}
\ln \delta (\theta)= 2 \psi(1) - \psi(\theta) -\psi(1-\theta)
\ .
\end{equation}

Therefore, even with $s=t=0$, we find a convergent integral.
Its behavior near $x=0$ is
\begin{equation} \label{proper}
\sim \pi^{3/2} \int_0 {dx\over x} [\ln (1/x)]^{-3/2}
\end{equation}
The physical reason for the absence of an IR divergence at
vanishing momentum is the following. Even though there is a
massless intermediate state from the untwisted sector, the special
kinematics of this problem gives no IR divergence. For example, if
the $s$-channel intermediate state is a massless $6'-6'$ string,
which is a gauge boson, then it can carry arbitrary momentum $q$
along the directions of the $D6'$-brane orthogonal to the intersection.
These three components of momentum have to be integrated over:
\begin{equation}
\int d^3 q \int_0 dx \,x^{\alpha' q^2 - \alpha' s - 1} = \pi^{3/2}
(\alpha')^{-3/2} \int_0 dx  x^{-1-\alpha' s} [\ln (1/x)]^{-3/2}\ ,
\end{equation}
which converges near $x=0$ even for $s=0$. Thus, going from
the effective field theory to the form of the string integrand near $x=0$,
we find the replacement
\begin{equation} \label{ftst}
\int {d^3 q\over q^2- s} \rightarrow
\pi^{3/2} (\alpha')^{-1/2}
\int_0 dx x^{-1-\alpha' s} [\ln (1/x)]^{-3/2}\ .
\end{equation}

This observation also allows us to normalize the 4-fermion
amplitude. If we consider the $s$-channel factorization on the
$6'-6'$ gauge boson, which is the leading term as $x\rightarrow 0$, then
we should find
\begin{equation} \label{unit}
A_4 (k_1,k_2,k_3,k_4) \rightarrow - i \int {d^7 k\over (2\pi)^7}
{\sum_{IJ\mu}A^{I \mu }_{\ J} (k_1,k_2, k) A^{J}_{\ I \mu } (k_3,
k_4, -k) \over k^2 - i\epsilon} \ ,
\end{equation}
where the 3-point function for emission of a gauge boson of
Chan-Paton indices $J,I$ is
\begin{equation}
A^{I \mu }_{\ J} (k_1,k_2, k_3) = i K \sqrt{g_s} (\alpha')^{3/4}
(2\pi)^4 \delta^4 (\sum_{i=1}^3 k_i) \bar u_1 \gamma^\mu u_2
\left(\tilde \lambda_{1 } \lambda_{2 }\right)^I{}_J \ .
\end{equation}
The index $\mu$ corresponds to the polarization of the
intermediate vector state. Since the gauge coupling on D6-branes
is given (according to \cite{Polch}, eqn. 13.3.25) by $g_{D6}^2 =
(2\pi)^4 g_s (\alpha')^{3/2}$, we have
\begin{equation}
K^2 = (2\pi)^4
\ .
\end{equation}

Performing the integral on the RHS of (\ref{unit}) with
the replacement (\ref{ftst}), we find
\begin{equation}
i \Tr (\tilde\lambda_1 \lambda_2 \tilde\lambda_3 \lambda_4)
K^2 g_s (\alpha') (2\pi)
\delta^4 (\sum_{i=1}^4 k_i)
\bar u_1 \gamma^\mu u_2 \bar u_3 \gamma_\mu u_4
\pi^{3/2}
\int_0 x^{-1-\alpha' s} [\ln (1/x)]^{-3/2}\ .
\end{equation}
Equating this to the contribution to $A_4$, given in eqn.
(\ref{ampl}), from the $s$-channel $6'-6'$ massless intermediate state,
\begin{equation}
i\Tr (\tilde\lambda_1 \lambda_2 \tilde\lambda_3 \lambda_4) C (2\pi)^4
\delta^4 (\sum_{i=1}^4 k_i)
\bar u_1 \gamma^\mu u_2 \bar u_3 \gamma_\mu u_4
\pi^{3/2}
\int_0 x^{-1-\alpha' s} [\ln (1/x)]^{-3/2}\ ,
\end{equation}
we find that
\begin{equation}
C= (2\pi)^{-3} K^2 g_s \alpha' = 2\pi g_s \alpha'
\ .
\end{equation}

The low-energy limit of the 4-fermion amplitude is hence,
\begin{equation}\eqalign{
A_4 (k_1,k_2,k_3,k_4) = &i (2\pi g_s \alpha')
I(\theta_1,\theta_2,\theta_3) (2\pi)^4 \delta^4 (\sum_{i=1}^4 k_i)
\bar u_1 \gamma^\mu u_2 \bar u_3 \gamma_\mu u_4 \cr &
\Tr\left(\tilde\lambda_1 \lambda_2 \tilde\lambda_3
\lambda_4+\tilde\lambda_1 \lambda_4 \tilde\lambda_3
\lambda_2\right)\ ,\cr}
\end{equation}
where
\begin{equation}
I(\theta_1,\theta_2,\theta_3)=
\int_0^1 {dx\over x(1-x)}
\prod_{i=1}^3 \sqrt{\sin (\pi \theta_i)}
[ F(\theta_i , 1-\theta_i; 1; x)
F(\theta_i , 1-\theta_i; 1; 1-x) ]^{-1/2}
\ .
\end{equation}

This has been derived for intersections of two stacks of branes
without any orientifolding.  Thus, in an intersection of a stack
of $N$ $D6$-branes with a stack of $M$ $D6'$-branes, we would have
a $U(N)\times U(M)$ gauge theory with the fermions supported at
the intersection points transforming as bifundamentals, and the
above four-fermi interactions.  To get an $SU(5)$ theory with
fields transforming in the ${\bf 10}$, we should consider the case
that $N=M=5$ and the two stacks are exchanged by an orientifolding
operation $\Omega R$ and intersect on the orientifold
plane ($\Omega$ is the worldsheet parity; $R$ acts by
complex conjugating all three
complex coordinates: $z_i \rightarrow \bar z_i$,
$i=1,2,3$). Moreover, we
pick an orientifolding operation  that projects onto fermions
whose wavefunctions $\lambda_i$ and $\tilde \lambda_j$ transform
as antisymmetric, rather than symmetric, tensors of $SU(5)$. Tree
level amplitudes in the orientifold theory are obtained by
computing tree level amplitudes on the covering space for states
that are invariant under the orientifolding projection, and then
dividing by 2.  The ${\bf 10}^2 {\bar{\bf 10}}{}^2$ interaction in
an orientifold theory is thus derived from
\begin{equation}\label{orientamp}\eqalign{
A_4 (k_1,k_2,k_3,k_4) = &i (\pi g_s \alpha')
I(\theta_1,\theta_2,\theta_3) (2\pi)^4 \delta^4 (\sum_{i=1}^4 k_i)
\bar u_1 \gamma^\mu u_2 \bar u_3 \gamma_\mu u_4 \cr &
\Tr\left(\tilde\lambda_1 \lambda_2 \tilde\lambda_3
\lambda_4+\tilde\lambda_1 \lambda_4 \tilde\lambda_3
\lambda_2\right)\ .\cr}
\end{equation}

\bigskip\noindent{\it Evaluation Of $I$}

Now let us discuss the numerical evaluation of the integral $I$.
After writing
\begin{equation}
{1\over x(1-x)}=
{1\over x}+
{1\over 1- x}
\ ,\end{equation}
noting that the
two terms contribute equally, and setting $x=e^{-t}$ to evaluate
the contribution of the first term, we can write $I$ as
\begin{equation}
I(\theta_1,\theta_2,\theta_3)= 2
\int_0^\infty dt
\prod_{i=1}^3 \sqrt{\sin (\pi \theta_i)}
[ F(\theta_i , 1-\theta_i; 1; e^{-t})
F(\theta_i , 1-\theta_i; 1; 1-e^{-t} ) ]^{-1/2}
\ .
\end{equation}
Let us try to evaluate the integral on the RHS numerically for the
most symmetric choice of rotation angles:
$\theta_1=\theta_2=\theta_3=2/3$. It turns out that for
sufficiently large $t$, it is hard to maintain numerical precision
in evaluating the hypergeometric functions. To deal with this
problem, we will break up the integral into the range from $0$ to
$T$ and from $T$ to $\infty$. In the first region we use
Mathematica to evaluate it numerically, while in the second we may
use the asymptotics (\ref{asymptotics}) to replace the integral by
\begin{equation}
2\pi^{3/2} \int_T^\infty
dt (t+ 3\ln 3)^{-3/2} =
4 \pi^{3/2}
(T + 3\ln 3)^{-1/2}\ ,
\end{equation}
where we used $\ln\delta (2/3) = 3\ln 3$. In practice, the sum of
the two is insensitive to $T$ in a certain range; we have checked
this for $T$ between 15 and 25. We find
\begin{equation}
I(2/3,2/3,2/3)\approx 11.52\ .
\end{equation}
It is interesting that a large number $11.52$ emerges from an
explicit string calculation.

Now cosider a generalization to $\theta_1=\theta_2=\theta$ and
$\theta_3= 2- 2\theta$.
For the purpose of numerical evaluation, we approximate
$I(\theta,\theta,2-2\theta)$ by
\begin{equation} \label{double}
2 \sin (\pi \theta)
\sqrt{-\sin (2\pi \theta)}
\int_0^T dt
{[ F(2-2\theta , 2\theta-1; 1; e^{-t})
F(2-2\theta , 2\theta-1; 1; 1-e^{-t} ) ]^{-1/2}
\over
F(\theta , 1-\theta; 1; e^{-t})
F(\theta , 1-\theta; 1; 1-e^{-t} )}
\end{equation}
$$
+2\pi^{3/2} \int_T^\infty {dt\over (t+\ln\delta (\theta)) \sqrt {t+
\ln\delta (2- 2\theta )} }\ .
$$
It is interesting to ask how $I$ behaves when one of the rotation
angles becomes small. This can be determined from (\ref{double})
by letting $\theta$ approach $1/2$ from above, so that
$\theta_3=2-2 \theta$ corresponds to rotation by close to $\pi$. It turns out that $I$ decreases
rather slowly from its maximum value at $\theta=2/3$. For example,
$I(0.55,0.55,0.9)\approx 9.89$ and $I(0.51,0.51,0.98)\approx
6.69$. Even when all three rotation angles become
effectively small, $I$ does
not fall off very rapidly. Using (\ref{double}), we find
$I(0.9, 0.9, 0.2)\approx 7.675$; $I(0.95, 0.95, 0.1)\approx 5.505$;
$I(0.99, 0.99, 0.02)\approx 2.47$. We conclude that for a broad
range of angles the quantity $I$ lies in the range $7 - 11.5$.

\section{Comparison To Four-Dimensional GUT's}

According to \cite{Polch}, eqns. (12.1.10b) and (13.3.24), the
gravitational action for Type IIA superstrings is
\begin{equation}
{1\over (2\pi)^7\alpha'^4}\int d^{10}x\sqrt{- G}e^{-2\Phi} R.
\end{equation}
The string coupling constant is $g_{s}=e^\Phi$. After reduction to
four dimensions on a six-manifold $X$  of volume $V_6$, the
gravitational action in four dimensions is
\begin{equation}
{V_6\over (2\pi)^7 g_{s}^2\alpha'^4}\int \sqrt{-g}R.
\end{equation}
As the coefficient of $R$ in four-dimensional General Relativity
is
conventionally $(16\pi G_N)^{-1}$, with $G_N$ Newton's constant, we
have
\begin{equation}\label{firstone}
g_s^2\alpha'^4={8 V_6G_N\over (2\pi)^6}. \end{equation}

The gauge coupling $g_{D6}$ of gauge fields on a $D6$-brane is
defined by saying that the effective action for the gauge fields
is
\begin{equation}
{1\over 4g_{D6}^2}\int d^7x\sqrt{g_7}\Tr\,F_{ij}F^{ij},
\end{equation}
where $F_{ij}$ is the Yang-Mills field strength and $\Tr$ is the
trace in the fundamental representation of $U(N)$. If we take the
$D6$-brane worldvolume to be ${\bf R}^4\times Q$, where $Q$ is a
compact three-manifold of volume $V_Q$, then the action in four
dimensions becomes
\begin{equation}
{V_Q\over 4g_{D6}^2}\int d^4x \Tr F_{ij}F^{ij}. \end{equation}
However, before comparing to four-dimensional GUT's, we must
recall that it is conventional to expand the gauge fields as
$A_i=\sum_a Q_a A_i^a$, where $\Tr\,Q_aQ_b={1\over 2}\delta_{ab}$.
Similarly, one conventionally expands $F_{ij}=\sum_aF_{ij}^aQ_a$.
If this is done, the action becomes
\begin{equation}
{V_Q\over 8g_{D6}^2}\int d^4x \sum_a F_{ij}^a F^{ija}.
\end{equation}
The GUT action is conventionally written
\begin{equation}
{1\over 4g_{GUT}^2}\int d^4x \sum_a F_{ij}^a F^{ija},
\end{equation}
where $g_{GUT}$ is the GUT coupling.  Hence, we have
$g_{GUT}^2=2g_{D6}^2/V_Q$.  Since
$g_{D6}^2=(2\pi)^4g_s{\alpha'}^{3/2}$ according to eqn. (13.3.25)
of \cite{Polch}, we have
\begin{equation}\label{ggut}
g_{GUT}^2={2(2\pi)^4g_s{\alpha'}^{3/2}\over V_Q}.
\end{equation}

The volume $V_Q$ enters in the running of $SU(3)\times SU(2)\times
U(1)$ standard model gauge couplings from very high energies down
to the energies of accelerators.  Roughly speaking, $V_Q^{-1/3}$
plays the role of $M_{GUT}$, the mass scale of unification, in a
four-dimensional $GUT$ theory.  To find the precise relation
between them, one must compute the one-loop threshold corrections
to gauge couplings.  This was done in \cite{FW} for $M$-theory on
a manifold of $G_2$ holonomy.  The one-loop threshold corrections
are the same in Type IIA as in $M$-theory, since they come from
Kaluza-Klein harmonics on $Q$ that are the same in the two
theories.  So we can borrow the result of \cite{FW}. According to
that result (see eqn. (3.30) of \cite{FW}), the precise relation
between $M_{GUT}$ understood as the mass at which the low energy
coupling constants appear to unify and $V_Q$ is
\begin{equation}\label{vsubq} V_Q={L(Q)\over
M_{GUT}^3},\end{equation}
 where $L(Q)$ is a certain  topological
invariant of $Q$ (together with Wilson lines on $Q$ that break the
GUT symmetry to the standard model) that is known as the
Reidemeister or Ray-Singer torsion.  $L(Q)$ depends on a model,
but is readily computable in a given model.  For example, in a
model considered in \cite{FW} in which $Q$ is a lens space ${\bf
S}^3/{\bf Z}_q$ for some positive integer $q$ (which must be prime
to 5 if we want an $SU(5)$ model), and the Wilson line on $Q$ has
eigenvalues $\exp(2\pi i\delta_i/q)$ with
$\delta_i=(2w,2w,2w,-3w,-3w)$ for some integer $w$ prime to $q$,
one has
\begin{equation}\label{thresh}
L(Q)=4q\sin^2(5\pi w/q).
\end{equation}
For a slightly more general lens space, also described in
\cite{FW}, whose definition depends on another integer $j$, this
is replaced by $L(Q)=4q|\sin(5\pi w/q)\sin(5\pi jw/q)|$. The
factor $L(Q)$ can be important in determining the proton lifetime,
as will become clear.  For example, for the minimal choice that
leads to the standard model gauge symmetry, which is $q=2$, $w=1$,
we have $L(Q)=8$, and the threshold correction will prove to
enhance the proton decay rate significantly. It is possible for
lens spaces to make $L(Q)<1$, but only if $q$ is extremely large.

Using  (\ref{thresh}), we can express (\ref{ggut}) in the form
\begin{equation}
\alpha_{GUT}=(2\pi)^3L(Q)^{-1}g_s{\alpha'}^{3/2}M_{GUT}^3,
\end{equation}
where as usual $\alpha_{GUT}=g_{GUT}^2/4\pi$.  Alternatively,
\begin{equation}\label{secondone}
g_s^2{\alpha'}^3={\alpha_{GUT}^2L(Q)^2\over (2\pi)^6M_{GUT}^6}.
\end{equation}

Ideally, we would like to compute all of the parameters in the
string compactification in terms of the quantities
$G_N=1/M_{Pl}^2$, $M_{GUT}$, and $\alpha_{GUT}$, about which we
have at least some experimental knowledge. The Planck mass is
well-known ($M_{Pl}\cong 1.2\times 10^{19}$ GeV), but $M_{GUT}$
and $\alpha_{GUT}$ are somewhat model-dependent. The most commonly
cited values based on extrapolation from low energy data are
$M_{GUT}\cong 2\times 10^{16}$ GeV, $\alpha_{GUT}\cong .04$.
Unfortunately, the string theory really depends on four unknowns
$V_6$, $V_Q$, $g_s$, and $\alpha'$.

To parametrize our ignorance, we may introduce the dimensionless
quantity
\begin{equation}
\lambda={V_6\over V_Q^2},
\end{equation}
which is of order one if $X$ is fairly isotropic, and solve for
the other stringy parameters in terms of $G_N$, $M_{GUT}$,
$\alpha_{GUT}$, and $\lambda$. For $V_Q$, this has already been
done in (\ref{vsubq}).  Dividing (\ref{firstone}) by
(\ref{secondone}), we get
\begin{equation}\label{alphaprime}
\alpha'={8\lambda G_N\over\alpha_{GUT}^2}.
\end{equation}
We can similarly solve for $g_s$,
\begin{equation}\label{gst}
g_s={\alpha_{GUT}^4 L(Q)\over
2^{9/2}(2\pi)^3M_{GUT}^3G_N^{3/2}\lambda^{3/2}}.
\end{equation}

The factor about which we have the least
intuition is $\lambda$, which comes from a ratio of volumes.  To
try to get some intuition about the possible values of $\lambda$,
and also about whether the model makes sense, let us examine
quantitatively the formula (\ref{gst}) for $g_{s}$:
\begin{equation} \label{standval}
g_{s}=0.1\left({\alpha_{GUT}\over .04}\right)^4
\left({2\times 10^{16}\,{\rm GeV}\over M_{GUT}}\right)^3
{L(Q)\over \lambda^{3/2}}.
\end{equation}
Our calculation does not make much sense if $g_{s}\gg 1$, since then
we should do the computation in $M$-theory (leading back to the
discussion in section 5 of \cite{FW}) rather than in Type IIA
superstring theory.  If $g_{s}\ll 1$, our computation makes sense,
but it is implausible for the vacuum to be stabilized after
supersymmetry breaking in a way that would enable the world as we
see it to exist.  So the discussion makes most sense if $g_{s}$ is
relatively close to 1.  We note that happily (\ref{standval}) is
compatible with having $g_{s}$ near 1 while the GUT parameters
have their usual values, $L(Q)=8$, and $\lambda$ is not too far
from 1.

In view of the preceding discussion,
it is perhaps more useful to use (\ref{standval}) to solve
for $\lambda$, parametrizing our ignorance via the unknown value
of $g_{s}$, which we expect to be near 1:
\begin{equation}
\lambda=\alpha_{GUT}^{8/3}
{ M_{Pl}^2 \over M_{GUT}^2 }
{L(Q)^{2/3}\over 8 (2\pi)^2 g_{s}^{2/3}}.
\end{equation}
Now (\ref{alphaprime}) becomes
\begin{equation}
\alpha' =
{ \alpha_{GUT}^{2/3} L(Q)^{2/3}\over (2\pi)^2 g_{s}^{2/3}
M_{GUT}^2 }
\ .
\end{equation}
Finally, the amplitude for proton decay is, from
(\ref{orientamp}),
\begin{equation}\label{stringamp}
A_{st}=\pi\alpha'g_s I(\theta_1,\theta_2,\theta_3) = {
\alpha_{GUT}^{2/3} L(Q)^{2/3} g_{s}^{1/3}
I(\theta_1,\theta_2,\theta_3) \over 4\pi M_{GUT}^2 } \
,\end{equation} where $I$ is the integral discussed in the last
section. It is interesting to compare this result with the
M-theory estimate of \cite{FW}:
\begin{equation}
A_{M}\sim  { \alpha_{GUT}^{2/3} L(Q)^{2/3}
\over M_{GUT}^2 }
\ .\end{equation}
We find the same scaling with $\alpha_{GUT}$, $L(Q)$
and $M_{GUT}$ as in M-theory. If we keep these parameters and
the rotation angles
$\theta_i$ fixed, then we may study the amplitude as a function of $g_s$.
String theory indicates that this function behaves as
$g_s^{1/3}$ for small $g_s$; M-theory tells us that it approaches a constant
for large $g_s$.
Strictly speaking, (\ref{stringamp}) is
reliable for small $g_s$, but we will take it as a rough
approximation for $g_s$ of order $1$.

\bigskip\noindent
{\it Analogous Field Theory Amplitude}

Let us compare this to the analogous four-fermion amplitude in
four-dimensional GUT's. For simplicity, we will assume that (as in
$SU(5)$ models) all superheavy gauge bosons have the same mass
$M_X$.  $M_X$ is comparable to the unification scale $M_{GUT}$
inferred from the running of the low energy gauge couplings, but
differs from it, in general, by model-dependent factors. Exchange
of superheavy gauge bosons of mass $X$ gives an amplitude
\begin{equation}\label{gaugesum}
{g_{GUT}^2\over M_X^2}\sum_a\langle A_1|J_{\mu a}|A_2\rangle
\langle A_3| J^\mu_a|A_4\rangle,
\end{equation}
where we have labeled initial and final fermion states as
$A_1,A_2,A_3$, and $A_4$.  The sum runs over the superheavy gauge
bosons, but since the standard model gauge bosons contribute
baryon-number conserving interactions anyway and taking the sum
over all generators of the gauge group will lead to a simpler
formula, we will  do this.  (Standard model gauge boson masses are
near zero, not near $M_X$, so (\ref{gaugesum}) is not a good
approximation to their contribution.)  Note that, ignoring
worldsheet instantons, the string theory four-fermion amplitude is
$SU(5)$-invariant, since it comes from a local contribution that
does not see the symmetry breaking by Wilson lines, while the
four-fermion operator in GUT's is of course not $SU(5)$-invariant.

We first consider the case that the fermions transform as ${\bf
5}$'s of $SU(5)$, though the resulting ${\bf 5}^2{\bar{\bf
5}}{}^2$ amplitude is actually baryon number conserving.  To
get baryon nonconservation, we need to incorporate ${\bf
10}$'s as well, as we will do presently.  Also, to get a better
match to the local string theory construction with Chan-Paton
factors, we take the gauge group to be $U(5)$ instead of $SU(5)$;
the extra $U(1)$ gauge field does not violate baryon number
anyway.  (In string theory, the local construction at a particular
brane intersection point has $U(5)$ gauge symmetry, but globally
in realistic models the extra $U(1)$ is Higgsed by absorbing an RR
mode.)

Let us work out the group theory part of the matrix element in
eqn. (\ref{gaugesum}).  For this, we introduce  column and row
vectors $\alpha_i$ and $\bar \alpha_j$ for states transforming in
the ${\bf 5}$ or $\bar{\bf 5}$; the group theory factor of the
current-current matrix element becomes
\begin{equation}
\sum_a  \bar \alpha_1 T_a \alpha_2\,\, \bar \alpha_3 T^a \alpha_4,
\end{equation}
 where the
$T_a$ are traceless $5\times 5$ matrices that generate $U(5)$.  As
we noted earlier, the $T_a$ are conventionally normalized so that
$\Tr\,T_aT_b={1\over 2}\delta_{ab}$. It is straightforward to show
that
\begin{equation}\label{polyglo}
\sum_a  (\bar \alpha_1 T_a \alpha_2)\,\, (\bar \alpha_3 T^a
\alpha_4)={1\over 2}(\bar \alpha_1\alpha_4) (\bar \alpha_3
\alpha_2).
\end{equation}
The ${\bf 5}^2{\bf \bar 5}{}^2$ interaction in GUT's is hence
\begin{equation}\label{fiveamp}
{g_{GUT}^2\over 2M_X^2}\left[
(\bar\alpha_1\alpha_4)(\bar\alpha_3\alpha_2)+(\bar\alpha_1\alpha_2)
(\bar\alpha_3\alpha_4) \right],\end{equation} where we sum over
the two channels in which the gauge boson can be exchanged.

Now we move on to the ${\bf 10}^2{\bar{\bf 10}}{}^2$ interaction.
In order to make clear the origin of a certain factor of 2, we
will introduce the ${\bf 10}$ by a field theory version of the
orientifold that we used in string theory. The gauge group is
$U(5)\times U(5)$, with gauge fields $A$, $A'$ and the fermions
consist of fields $\psi,\psi'$ transforming as $({\bf 5},{\bf
1})+({\bf 1},\bar{\bf 5})$ plus a field $S$ transforming as $({\bf
5},\bar{\bf 5})$. The Lagrangian is invariant under an
``orientifolding''  symmetry $\Theta$ that exchanges $A$ with
$(A')^T$, where $(A')^T$ is the transpose (or equivalently, as
$A'$ is hermitian, the complex conjugate) of $A'$, breaking
$U(5)\times U(5)$ to a diagonal $U(5)$. $\Theta$ exchanges $\psi$
and $\psi'$, leaving a ${\bf 5}$ of the unbroken $U(5)$, and we
take it to act on $S$ with a suitable sign such that the invariant
modes in $S$ transform as the ${\bf 10}$ of the diagonal $U(5)$,
corresponding to an antisymmetric $5\times 5$ matrix $S^{ij}$.

The kinetic energy is taken to be a general $\Theta$-invariant
expression:
\begin{equation}\label{omigo}
I={1\over 4g_{GUT}^2}\Tr (F(A)^2+F(A')^2) +\bar \psi i\gamma\cdot D
\psi +\bar\psi' i\gamma\cdot D \psi' +\bar Si\gamma\cdot D S.
\end{equation}
where sums over all indices of $\psi,$ $\psi'$, and $S$ are
understood. Let us work out the $S^2\bar S{}^2$ interaction prior
to orientifolding.  It comes from exchanges of  gauge bosons in
the two $U(5)$ groups. In working out the contribution from
exchange of a gauge boson in {\it either} $U(5)$ factor, the
indices of $S$ that  transform under the other $U(5)$ are
spectators. The amplitude due to exchange of either $U(5)$ is thus
just like (\ref{fiveamp}), except that when we include the
spectator index, the wavefunctions all are $5\times 5$ matrices
$\tilde \lambda_i$ and $\lambda_j$ rather than row and column
vectors, and the pairing of row and column vectors is replaced by
a trace.  The amplitude due to exchange of either $U(5)$ is hence
\begin{equation}\label{secondamp}
{g_{GUT}^2\over 2M_{GUT}^2} \bar u_1 \gamma^\mu u_2 \bar u_3 \gamma_\mu u_4
\Tr(\tilde \lambda_1
\lambda_2\tilde\lambda_3\lambda_4+\tilde\lambda_1\lambda_4\tilde\lambda_3\lambda_2).
\end{equation}
The total amplitude due to exchange of a gauge boson in one or the
other $U(5)$ is hence
\begin{equation}\label{thirdamp}
{g_{GUT}^2\over M_{GUT}^2} \bar u_1 \gamma^\mu u_2 \bar u_3 \gamma_\mu u_4
\Tr(\tilde \lambda_1
\lambda_2\tilde\lambda_3\lambda_4+\tilde\lambda_1\lambda_4\tilde\lambda_3\lambda_2).
\end{equation}

Orientifolding is carried out by restricting to $\Theta$-invariant
states and dividing the action by 2.  As a result, we can drop
$A'$ and $\psi'$, but we must divide the $S$ kinetic energy by 2.
The resulting kinetic energy is then
\begin{equation}\label{orbkin}
I'={1\over 4 g_{GUT}^2}F(A)^2+ \bar\psi i\gamma\cdot D\psi
+{1\over 2}\sum_{ij}\bar S_{ij} i\gamma\cdot D S^{ij}.
\end{equation}
The $S$ kinetic energy has been divided by $2$, but is still
canonically normalized in the following sense.  $S$ is an
antisymmetric matrix, so for example has a $2\times 2$ block
looking like
\begin{equation}\label{twobytwo}
\left(\matrix{ 0 & e^+ \cr -e^+ & 0\cr}\right).
\end{equation}
The kinetic energy is canonically normalized for fields like $e^+$
appearing above the diagonal in $S$.

As for the ${\bf 10}^2{\overline{\bf 10}}{}^2$ interaction in the
orientifold theory, we can borrow it from (\ref{thirdamp}).  We
merely have to take the wavefunctions $\lambda_i$ and $\tilde
\lambda_j$ to be invariant under $\Theta$, which means that they
are $5\times 5$ antisymmetric matrices.  Also, we have to divide
by 2, just as we divided the $\Theta$-invariant part of the
classical action by 2 to get (\ref{orbkin}).  So the ${\bf
10}^2\bar{\bf 10}^2$ interaction is finally
\begin{equation}\label{fourthamp}
{g_{GUT}^2\over 2M_{GUT}^2} \bar u_1 \gamma^\mu u_2 \bar u_3 \gamma_\mu u_4
\Tr(\tilde \lambda_1
\lambda_2\tilde\lambda_3\lambda_4+\tilde\lambda_1\lambda_4\tilde\lambda_3\lambda_2).
\end{equation}

Of course, we could have derived this directly starting with the
kinetic energy (\ref{orbkin}).

\bigskip\noindent{\it Comparison}

 To compare string theory to
four-dimensional gauge theory, we therefore need merely compare
the string theory amplitude $\pi g_{s}\alpha'I$ from
(\ref{orientamp}) to the field theory amplitude
$g_{GUT}^2/2M_X^2=2\pi \alpha_{GUT}/M_X^2$ from (\ref{fourthamp}).
The comparison is thus
\begin{equation}
\pi  g_{s}\alpha'I\leftrightarrow 2\pi{\alpha_{GUT}\over M_X^2}.
 \end{equation}
We call the left hand side the string amplitude $A_{st}$, and the
right hand side the corresponding GUT amplitude $A_{GUT}$.  Using
(\ref{stringamp}), we find the ratio to be

\begin{equation}\label{polydon}
{A_{st}\over A_{GUT}}={L(Q)^{2/3}I\over 8\pi^2} {g_{s}^{1/3}\over
\alpha_{GUT}^{1/3}} {M_X^2\over M_{GUT}^2}
\end{equation}
For other parameters fixed, this ratio scales as
$1/\alpha_{GUT}^{1/3}$, which is the same as in the M-theory
calculation of \cite{FW}. The negative power of $\alpha_{GUT}$
reflects the fact that in this type of model, proton decay is a
stringy effect, rather than being dominated by exchange of a
single Kaluza-Klein mode.  Exploring a possible enhancement due to
this factor was the motivation for the present paper, but the
factor of $8\pi^2$ in the denominator is clearly unfavorable. We
can rewrite (\ref{polydon}) as
\begin{equation}\label{ampcompar}
{A_{st}\over A_{GUT}} \approx 0.037 L(Q)^{2/3}I
g_{s}^{1/3}\left(.04\over \alpha_{GUT}\right)^{1/3}{M_X^2\over
M_{GUT}^2}.
\end{equation}
Happily, $g_{s}$, of which we can only say that we expect it to be
more or less close to 1, is here raised to a relatively small
power, helping to reduce the uncertainty.

Using the most commonly cited values $M_{GUT}\cong 2\times
10^{16}$ GeV, $\alpha_{GUT}\cong .04$, a recent evaluation of
 the proton lifetime in
four-dimensional SUSY $SU(5)$ due to gauge boson exchange gave the
value $1.6\times 10^{36}$ years if $M_X=M_{GUT}$ \cite{hisano} (we
have taken eqn. (17) of this paper with $M_X=2\times 10^{16}$
GeV). If $M_X$ or $\alpha_{GUT}$ is changed, the proton lifetime
scales as:
\begin{equation}\label{plife}
\tau_p=1.6\times 10^{36}{\rm years}\left({.04\over
\alpha_{GUT}}\right)^2\left(M_X\over 2\times 10^{16}\,{\rm
GeV}\right)^4.
\end{equation}
 Using modified
values of $\alpha_{GUT}$ and $M_{GUT}$ is natural in the present
discussion if doublet-triplet splitting is accomplished with
discrete symmetries as in \cite{witten}, since this necessitates
extra matter fields (in complete $SU(5)$ multiplets) surviving to
energies below the GUT scale. For example, in \cite{pati}, in
models with such additional multiplets, it was found plausible to
have values such as $\alpha_{GUT}=.2$, $M_{GUT}=8\times 10^{16}$
GeV. The net effect of these changes is to roughly double the
proton lifetime.

To obtain the proton lifetime $\tau_{p,st}$ in this class of
string theories from (\ref{plife}), we multiply by
$(A_{GUT}/A_{st})^2$  and replace the prefactor 1.6 by 2 for a
reason explained below. Thus:
\begin{equation}\label{stringlife}
\tau_{p,st}=2\times 10^{36}\,{\rm
years}\left(.037L(Q)^{2/3}Ig_s^{1/3}\right)^{-2}\left(.04\over
\alpha_{GUT}\right)^{4/3}\left(M_{GUT}\over 2\times 10^{16}\,{\rm
GeV}\right)^4.
\end{equation}

A proton lifetime of order $10^{36}$ years due to gauge boson
exchange is considered unobservably small for the foreseeable
future; the present experimental bound on $p\to \pi^0e^+$ is about
$4.4\times 10^{33}$ years (for example, see \cite{ganezer}), while
a next generation detector (described for example in \cite{jung})
may reach a limit close to $10^{35}$ years.  However, it is clear
that a relatively modest enhancement in the proton decay amplitude
might save the day.  Looking at (\ref{ampcompar}), we see a few
large and small factors that tend to cancel. This formula contains
the small overall factor $0.037$, along with the factor $I$, which
we have found to be roughly $7-11$ for plausible values of the
angles, and the threshold factor $L(Q)^{2/3}$, where $L(Q)$ need
not be large but is in fact equal to 8 for the minimal lens space
with fundamental group ${\bf Z}_2$, and a little larger for many
of the lens spaces with other small fundamental groups.  Combining
these factors,  the best we can say is that in a model based on
intersecting $D6$-branes, rather as in four-dimensional GUT's, the
proton lifetime due to dimension six operators is likely to be
close to $10^{36}$ years.

Note that one major source of uncertainty in GUT's is absent here:
the proton lifetime directly involves the scale $M_{GUT}$ that can
be probed using low energy data, rather than the heavy gauge boson
mass $M_X$ whose relation to $M_{GUT}$ is model-dependent.

What we have evaluated in this paper is a local, stringy
contribution to proton decay in a certain class of models based on
intersecting Type II $D6$-branes. If the compactification volume
is even slightly larger than the string scale, so that worldsheet
instanton effects are at least slightly suppressed, what we have
computed is likely to be the dominant contribution (from dimension
six operators) in this class of models. The result we have
obtained has an anomalous power of $\alpha_{GUT}$ because it is a
short distance stringy effect. As in \cite{FW}, this contribution
yields only a ${\bf 10}^2{\bf{\overline {10}}}{}^2$ interaction,
which contributes to $p\to\pi^0e_L^+$.  In four-dimensional GUT's,
there is also a ${\bf 10}\,{\bf {\overline {10}}} \,{\bf 5}\,{\bf
\overline 5}$ contribution, leading to $p\to \pi^0e^+_R$. With the
assumptions made in \cite{hisano}, the ratio of $p\to \pi^0e^+_R$
to the total is $y=1/(1+(1+|V_{ud}|^2)^2 )$, where $V_{ud}\cong 1$
is a CKM matrix element. Hence in comparing the stringy proton
decay rate to that in field theory, we should include a factor of
$1/(1-y)$ in the proton lifetime in the string model to account
for the missing $p\to \pi^0e^+_L$.  We included this factor in
taking the prefactor in (\ref{stringlife}) to be 2.

The last paragraph has been formulated loosely; with different
assumptions about the flavor structure at the GUT scale, the
$\pi^0$ could be a $K^0$ and the $e^+$ could be a $\mu^+$ (in
which case the lepton polarization in the final state would be
measurable). In either the GUT theory or the string model, the
proton lifetime could be increased further by unfavorable
assumptions about flavor structure (mixing with the third
generation, for instance, or mixing with new GUT-scale fermions).
At any rate, the assertion of this class of models that proton
decay is caused mainly by ${\bf 10}^2{\bar {\bf 10}}^2$
interactions is testable in principle.

\section*{Acknowledgments}
We thank M. Cvetic and S. Raby for useful discussions.
I.R.K. gratefully acknowledges the hospitality of the
George P. \& Cynthia W. Mitchell Institute for Fundamental
Physics at Texas A\& M University, where some of this work was
carried out.
This material is based upon
work supported in part by the National Science Foundation Grants
No. PHY-9802484 and PHY-0070928. Any opinions, findings, and
conclusions or recommendations expressed in this material are
those of the authors and do not necessarily reflect the views of
the National Science Foundation.

\begingroup\raggedright\endgroup


\end{document}